\documentclass[12pt]{article}
\usepackage{epsfig}
\usepackage{amsfonts}
\usepackage{amsopn}
\usepackage{amsmath}
\usepackage{verbatim,amsthm}

\DeclareMathOperator{\ad}{ad}
\DeclareMathOperator{\bt}{\boldsymbol{\theta}}

\DeclareMathOperator{\grad}{grad}

\DeclareMathOperator{\Spin}{\textrm{Spin}}
\DeclareMathOperator{\SO}{\textrm{SO}}
\DeclareMathOperator{\SU}{\textrm{SU}}

\DeclareMathOperator{\transp}{\textrm{T}}
\DeclareMathAlphabet{\mathpzc}{OT1}{pzc}{m}{it}

\title{Spin(9) Average of SU(N) Matrix Models\\I. Hamiltonian}

\author{
Jens Hoppe$^a$ \thanks{e-mail:  hoppe@math.kth.se} \ ,  \
Douglas Lundholm$^a$  \thanks{e-mail: dogge@math.kth.se} \\
and Maciej Trzetrzelewski$^{a,b}$ \thanks{e-mail: 33lewski@th.if.uj.edu.pl}
}

\date{}

\begin{document}

\maketitle

\begin{abstract}
We apply a method of group averaging to states and operators 
appearing in (truncations of) the $\Spin(9) \times \SU(N)$ invariant matrix models.
We find that there is an exact correspondence between the standard supersymmetric
Hamiltonian and the $\Spin(9)$ average of a relatively simple lower-dimensional model.
\end{abstract}

\vfill

\parbox[c]{\textwidth}{ \footnotesize
$^a$  Department of Mathematics,
Royal Institute of Technology, \\
KTH, 100 44 Stockholm, 
Sweden \\ \\
$^b$ Institute of Physics,
Jagiellonian University, \\
Reymonta 4, 30-059 Krak\'ow,
Poland
}

\newpage

\setlength\arraycolsep{2pt}
\def\arraystretch{1.5}

\section{Introduction}

Due to its relevance to M-theory, reduced Yang-Mills theory, and membrane theory, 
considerable effort has been put into investigating the structure of
Spin(9) $\times$ SU(N) invariant supersymmetric matrix models
(see e.g. \cite{review} for a review).
Despite this, a concrete knowledge of the conjectured zero-energy eigenfunction
of the Hamiltonian $H$ is still lacking.

In \cite{Hoppe3} a certain truncation of the Spin(9) invariant model was introduced, 
based on a coordinate split of $\mathbb{R}^9$ into $\mathbb{R}^7 \times \mathbb{R}^2$.
The corresponding Hamiltonian $H_D = \{Q_D,Q_D^\dagger\}$,
which is essentially just a set of supersymmetric harmonic oscillators,
can be interpreted as a two-dimensional supersymmetric SU(N) matrix model 
with a seven-dimensional space of parameters.
Recently, a deformation 
of the standard matrix model -- based on the same coordinate split --
was considered that
produces a $G_2 \times U(1)$ invariant supersymmetric Hamiltonian
$\tilde{H}$ in which $H_D$ plays a central role \cite{octonionic}. 
The explicit knowledge of
the structure of $H_D$ and its eigenfunctions made it possible to prove
in a straightforward manner that $\tilde{H}$ 
and $H$ have similar spectra.

In this paper we calculate the Spin(9) average of the truncated Hamiltonian $H_D$
and find that it is essentially equal to the full supersymmetric Hamiltonian $H$.
The correspondence is made exact by a slight modification of $H_D$.

Motivated by this result,
we also expect that the \emph{wavefunctions} obtained by averaging the
eigenfunctions of $H_D$ (or slight modifications of those) could be related
to the Spin(9) invariant eigenfunctions of $H$.
Calculating the average of such eigenstates, however, is a technically more difficult problem,
to be addressed in a forthcoming paper \cite{avg_state}.

\section{Group averaging}

First, let us define what we mean by group averaging 
(the notion is well-known in the literature; see 
e.g. \cite{Thiemann}
and references therein for a general approach and various applications).

Assume that we are given a unitary representation $U(g)$ of a compact Lie group $G$ acting
on a complex separable Hilbert space $\mathcal{H}$. Then, given any state 
$\Psi \in \mathcal{H}$ and linear operator $A$ acting on $\mathcal{H}$,
we define the corresponding $G$-averaged state $[\Psi\rangle_G$ resp. operator $[A]_G$ by
$$
	[\Psi\rangle_G := \int_{g \in G} U(g)\Psi \ d\mu(g)
$$
resp.
$$
	[A]_G := \int_{g \in G} U(g)AU(g)^{-1} \ d\mu(g),
$$
where $\mu$ denotes the unique \emph{normalized}
left- and right-invariant (Haar) measure on $G$.
Due to the translation invariance of $\mu$, $[\Psi\rangle_G$
will be invariant under the action of $U(g)$, and $[A]_G$ will commute with $U(g)$.

One can also extend the above definition to generalized (non-nor\-mal\-iz\-able) states, 
e.g. Schwartz distributions $\psi \in \mathcal{D}'$, by taking\footnote{Note
that the action of $G$ on distributions is given by $(U(g)\psi)(\phi) = \psi(U(g^{-1})\phi)$.}
$$
	\langle\psi]_G(\phi) := \int_{g \in G} \psi \left( U(g)\phi \right) \ d\mu(g)
$$
for any test function $\phi \in \mathcal{D} = C_0^\infty(\Omega)$.

\section{The model and its group actions}

We are interested in the supersymmetric matrix model described by the Hilbert space 
$$
	\mathcal{H} = L^2(\mathbb{R}^{9n}) \otimes \mathcal{F}, 
	\qquad \mathcal{F} = \mathop{\otimes}_{A=1}^{n} \mathcal{F}^{(A)} = \mathbb{C}^{2^{8n}}
$$
and the Hamiltonian
$$
    H = p_{sA}p_{sA} + \frac{1}{2}(f_{ABC}x_{sB}x_{tC})^2
    + \frac{i}{2} x_{sC} f_{ABC} \gamma^s_{\alpha \beta} \bt_{\alpha A}\bt_{\beta B}
	= -\Delta + V + H_F,
$$
where we sum over corresponding indices 
$s,t,\ldots=1,\ldots,9$, $A,B,\ldots=1,\ldots,n := N^2-1$, $\alpha,\beta,\ldots=1,\ldots,16$.
$\gamma^s$ generate (a matrix representation of)
the Clifford algebra over $\mathbb{R}^9$ acting irreducibly on $\mathbb{R}^{16}$,
while $\bt_{\alpha A}$ generate the Clifford algebra over 
$\mathbb{R}^{16} \otimes \mathbb{R}^n$, i.e. 
$\{ \bt_{\alpha A} , \bt_{\beta B} \} = 2\delta_{\alpha\beta}\delta_{AB}$,
acting irreducibly on $\mathcal{F}$.
The coordinates $x_{sA}$, canonically conjugate to $p_{sA} = -i\partial_{sA}$, 
comprise a set of 9 traceless hermitian matrices 
$(X_1,\ldots,X_9) = \boldsymbol{X} \in \mathbb{R}^9 \otimes \mathbb{R}^n$,
and we use the isomorphism $i \cdot \mathfrak{su}(N) \cong \mathbb{R}^n$
to map seamlessly between such a matrix $E$ and its coordinate representation
$e_A$ in a basis where the $\SU(N)$ structure constants $f_{ABC}$ are
totally antisymmetric.

$H$ is invariant under the action of $\SU(N)$, where the corresponding
representation on $\mathcal{H}$ is generated by the anti-hermitian operators
\begin{eqnarray*}
    \tilde{J}_A &=& iJ_A = if_{ABC}\Big( x_{sB}p_{sC} - \frac{i}{4}\bt_{\alpha B}\bt_{\alpha C} \Big) \\
	&=& \sum_{B<C} f_{ABC} \Big( x_{sB}\partial_{sC} - x_{sC}\partial_{sB} + \frac{1}{2} \bt_{\alpha B}\bt_{\alpha C} \Big)
	= \tilde{L}_A + \tilde{M}_A,
\end{eqnarray*}
with $\tilde{L}_A$ and $\tilde{M}_A$ generating the representation of 
$\mathfrak{su}(N) \hookrightarrow \mathfrak{so}(n)$ on 
$L^2(\mathbb{R}^{9n})$ and $\mathcal{F}$, respectively.

Furthermore, $H$ is also invariant under $\Spin(9)$, generated by
$$
	\tilde{J}_{st} = iJ_{st} 
	= i\Big( x_{sA}p_{tA} - x_{tA}p_{sA} - \frac{i}{8} \gamma^{st}_{\alpha\beta} \bt_{\alpha A}\bt_{\beta A} \Big)
	= \tilde{L}_{st} + \tilde{M}_{st},
$$
with
$$
	\tilde{L}_{st} = \sum_A \tilde{L}^{(A)}_{st} = x_{sA}\partial_{tA} - x_{tA}\partial_{sA}
$$
and 
$$
	\tilde{M}_{st} = \sum_A \tilde{M}^{(A)}_{st} = \sum_{\alpha<\beta} \frac{1}{2} [{\textstyle \frac{1}{2}\gamma^{st}}]_{\alpha\beta} \bt_{\alpha A}\bt_{\beta A}.
$$
Note that the spinor representation of $\Spin(9)$ is generated by 
$\frac{1}{2}\gamma^{st} := \frac{1}{4}[\gamma^{s},\gamma^{t}]$
acting by left multiplication on the Clifford algebra generated by the $\gamma$'s, 
i.e. left multiplication by the matrix 
$[\frac{1}{2}\gamma^{st}] \in \mathfrak{so}(16)$ 
acting on the spinor space $\mathbb{R}^{16}$.
This action is in turn represented on the Fock space $\mathcal{F}^{(A)} = \mathbb{C}^{2^8}$
by the spinor representation of $\mathfrak{spin}(16) = \frac{1}{2} \cdot \mathfrak{so}(16)$.

The full, exponentiated, action of 
$g = e^{\epsilon_{st}\frac{1}{2}\gamma^{st}} \in \Spin(9)$ on a state 
$\Psi \in \mathcal{H}$, i.e. a wavefunction
$\Psi: \mathbb{R}^9 \otimes \mathbb{R}^n \to \mathcal{F}$, is then given by
$$
	( U(g) \Psi )(\boldsymbol{X}) 
	= ( e^{\epsilon_{st} \tilde{J}_{st}} \Psi )(\boldsymbol{X})
	= e^{\epsilon_{st} \tilde{M}_{st}} \Psi(e^{\epsilon_{st} \tilde{L}_{st}} \boldsymbol{X})
	= \tilde{R}_g \Psi(R_{g^{-1}}(\boldsymbol{X})),
$$
where $\tilde{R}_g := e^{\epsilon_{st}\frac{1}{8}\gamma^{st}_{\alpha\beta} \bt_{\alpha A}\bt_{\beta A}}$ 
is the unitary representative of $g$ acting on $\mathcal{F}$,
and $R_g(\boldsymbol{x}) := g\boldsymbol{x}g^{-1}$ is the corresponding 
rotation $R_g \in \SO(9)$ acting on vectors $\boldsymbol{x} = x_t\gamma_t \in \mathbb{R}^9$ 
considered as grade-1 elements of the Clifford algebra.
This follows by considering the infinitesimal action on a function 
$f: \mathbb{R}^9 \to \mathbb{R}$, i.e. 
$(\tilde{L}_{st}f)(\boldsymbol{x}) = \grad f \cdot \tilde{L}_{st}(\boldsymbol{x})$,
and (using $\boldsymbol{x} \cdot \boldsymbol{y} = \frac{1}{2}\{\boldsymbol{x},\boldsymbol{y}\}$)
\begin{eqnarray*}
	\tilde{L}_{st}\boldsymbol{x} &=& (x_s\partial_t - x_t\partial_s)(x_u\gamma_u) = x_s\gamma_t - x_t\gamma_s \\
	&=& (\boldsymbol{x} \cdot \gamma_s) \gamma_t - (\boldsymbol{x} \cdot \gamma_t) \gamma_s 
	= - \frac{1}{2}[\gamma^{st},\boldsymbol{x}].
\end{eqnarray*}

Consider now an operator of the form
$$
	\mathcal{B} = [B(\boldsymbol{X})]_{\alpha\beta} \bt_{\alpha A} \wedge \bt_{\beta B}
$$
where $B(\boldsymbol{X})$ is a symmetric $16 \times 16$ matrix and $A \wedge B := \frac{1}{2}[A,B]$.
The infinitesimal action is
$$
	[\tilde{J}_{st} , \mathcal{B}] 
	= \left[ [\tilde{L}_{st}, B(\boldsymbol{X})] \right]_{\alpha\beta} \bt_{\alpha A} \wedge \bt_{\beta B} 
	\ +\ [B(\boldsymbol{X})]_{\alpha\beta} [\tilde{M}_{st}, \bt_{\alpha A} \wedge \bt_{\beta B}],
$$
which, using
\begin{eqnarray*}
	[\tilde{M}_{st}, \bt_{\alpha A} \wedge \bt_{\beta B}] 
	&=& \frac{1}{8} \gamma^{st}_{\alpha'\beta'} [\bt_{\alpha'C} \wedge \bt_{\beta'C}, \bt_{\alpha A} \wedge \bt_{\beta B}] \\
	&=& \frac{1}{2} (\gamma^{st}_{\epsilon \alpha} \bt_{\epsilon A} \wedge \bt_{\beta B} - \gamma^{st}_{\beta \epsilon} \bt_{\alpha A} \wedge \bt_{\epsilon B}),
\end{eqnarray*}
exponentiates to
\begin{eqnarray}
	e^{\epsilon_{st} \tilde{J}_{st}} \mathcal{B} e^{-\epsilon_{st} \tilde{J}_{st}}
	&=& \left[ e^{\frac{1}{2}\epsilon_{st}\gamma^{st}} B(e^{-\frac{1}{2}\epsilon_{st}\gamma^{st}} \boldsymbol{X} e^{\frac{1}{2}\epsilon_{st}\gamma^{st}}) e^{-\frac{1}{2}\epsilon_{st}\gamma^{st}} \right]_{\alpha\beta} \bt_{\alpha A} \wedge \bt_{\beta B}  \nonumber \\
	&=& [ gB(R_g^{\transp}(\boldsymbol{X}))g^{-1} ]_{\alpha\beta} \bt_{\alpha A} \wedge \bt_{\beta B}. \label{operator_action}
\end{eqnarray}

Regarding the supersymmetry of $H$, it is for the following sufficient
to know that there is a set of hermitian supercharge
operators $\mathcal{Q}_\alpha$ such that $H = \mathcal{Q}_\alpha^2$ on the subspace 
of $\SU(N)$ invariant states, $\mathcal{H}_{\textrm{phys}}$, 
which is the physical Hilbert space of the theory.

In order to arrive at a conventional Fock space formulation
of the model it is necessary to make certain choices which break the
explicit $\Spin(9)$ symmetry. After introducing a split
of the coordinates into 
$(x',z)$, with $x'= (x_{j=1,\ldots,7}) \in \mathbb{R}^{7n}$, $z_A := x_{8A} + ix_{9A}$,
and a representation of $\bt_{\alpha A}$ in terms of creation and annihilation operators $\lambda, \lambda^\dagger$,
together with a suitable representation of $\gamma^s$ (see e.g. Appendix A of \cite{octonionic}),
it is rather natural to single out
a certain part of the supercharges, 
resulting in a truncation of $H$ to 
the Hamiltonian \cite{Hoppe3,octonionic}
\begin{eqnarray*}
	H_D &=& -4 \bar{\partial}_z \cdot \partial_z + \bar{z} \cdot S(x') z + 2W(x')\lambda \lambda^\dagger 
	= -\Delta_{89} + V_D + W_D,
\end{eqnarray*}
where each of these terms will be explained in the next section.
This operator constitutes a set of $2n$ supersymmetric harmonic oscillators in $x_{8A}$ and $x_{9A}$
whose frequencies are the square root of the eigenvalues of the 
positive semidefinite matrix operator
$S(x') = \sum_{j=1}^7 \ad_{X_j} \circ \ad_{X_j}$. 
Thus, $H_D$ can be considered as acting on a smaller Hilbert space
over the {z}-coordinates,
$$
	\mathpzc{h} = L^2(\mathbb{R}^{2n}) \otimes \mathcal{F},
$$
with $x_j$ entering as parameters, 
and has with respect to 
$\mathpzc{h}$ the complete basis of eigenstates
\begin{equation} \label{eigenstates}
	\psi_{k,\sigma}(x',z) = \pi^{-\frac{n}{2}} (\det S(x'))^{\frac{1}{4}} H_k(x',z) e^{-\frac{1}{2}\bar{z}\cdot S(x')^{1/2} z} \xi_{x'}^\sigma.
\end{equation}
$H_k(x',z)$ denote products of normalized Hermite polynomials in 
$S(x')^{\frac{1}{4}}x_{8}$ and $S(x')^{\frac{1}{4}}x_{9}$, 
while $\xi_{x'}^\sigma \in \mathcal{F}$, $\sigma \in \{0,1\}^{8n}$, 
form the basis of eigenvectors of $W_D$
(see \cite{Hoppe3,octonionic} for details).

As pointed out in \cite{octonionic}, both $H_D$ and its nondegenerate eigenstates 
are SU(N) invariant (covariant) in the sense that they are unchanged
under the simultaneous action of $\SU(N)$ on $\mathpzc{h}$ and the parameters $x_j$.

\section{The averaged Hamiltonian}

We would like to apply group averaging w.r.t. $G = \Spin(9)$ to the 
truncated Hamiltonian $H_D$
and its $\mathpzc{h}$-eigenstates \eqref{eigenstates} 
(which are generalized states w.r.t. the full Hilbert space $\mathcal{H}$).

Note that averaging the supercharge $Q_D$ corresponding to $H_D$ 
gives zero in the same way that, 
for the supercharges $\mathcal{Q}_\alpha$ corresponding to $H$ and transforming like spinors,
$[\mathcal{Q}_\alpha]_G = [g_{\beta\alpha}\mathcal{Q}_\beta]_G = 0$, 
taking $g=-1$.

\subsection{Laplacian part}

The principal part of $H_D$ is the Laplace operator on $\mathbb{R}^{2} \otimes \mathbb{R}^n$,
$$
	\Delta_{89} = \Delta_8 + \Delta_9 = \partial_{8A}\partial_{8A} + \partial_{9A}\partial_{9A}.
$$
In order to average this operator,
consider first $x_1^2 = (\boldsymbol{x} \cdot \gamma_1)^2$ in $\mathbb{R}^d$, for which
\begin{eqnarray*}
	\lefteqn{
	[x_1^2]_{\Spin(d)} 
	= \int_{g \in \Spin(d)} U(g) (\boldsymbol{x} \cdot \gamma_1)^2 U(g)^{-1}\ d\mu(g) } \\
	&&= \int (R^{\transp}_g(\boldsymbol{x}) \cdot \gamma_1)^2\ d\mu(g)
	= \frac{1}{d} \sum_{j=1}^d \int (\boldsymbol{x} \cdot R_{gh_j}(\gamma_1))^2\ d\mu(g) \\
	&&= \frac{1}{d} \int \sum_{j=1}^d (\boldsymbol{x} \cdot R_{g}(\gamma_j))^2\ d\mu(g)
	= \frac{1}{d} \left[ |\boldsymbol{x}|^2 \right]_{\Spin(d)}
	= \frac{1}{d} |\boldsymbol{x}|^2,
\end{eqnarray*}
where we used the invariance of $\mu$ to insert $h_j \in \Spin(d)$ 
s.t. $R_{h_j}(\gamma_1) = \gamma_j$.
Analogously, one finds $[\partial_1^2]_{\Spin(d)} = \frac{1}{d} \Delta_{\mathbb{R}^d}$.
Hence,
$$
	[\Delta_{89}]_{\Spin(9)} = \frac{2}{9} \Delta_{\mathbb{R}^{9n}}.
$$

\subsection{Potential part}

Denoting the norm in $i \cdot \mathfrak{su}(N)$ by $\|\cdot\|$, so that for such a matrix
$E \leftrightarrow e_A$, $\|E\|^2 = e_Ae_A$,
we have
$$
	V_D 
	= \bar{z}_A S(x')_{AA'} z_{A'}
	= \bar{z}_A f_{ABC}x_{jB} f_{A'B'C}x_{jB'} z_{A'}
	= \sum_{\genfrac{}{}{0pt}{}{a=8,9}{j=1,\ldots,7}} \|[X_a, X_j]\|^2.
$$
Using that any pair $(\gamma_a,\gamma_j)$ of orthonormal vectors can be rotated
into any other orthonormal pair $(\gamma_s,\gamma_t) = (R_h(\gamma_a),R_h(\gamma_j))$ 
by some $R_{h}$, $h \in \Spin(9)$, we find
\begin{eqnarray*}
	\lefteqn{ \left[ \|[X_a,X_j]\|^2 \right]_G
	= \int_{G} U(g) \|[\boldsymbol{X} \cdot \gamma_a, \boldsymbol{X} \cdot \gamma_j]\|^2 U(g)^{-1}\ d\mu(g) }\\
	&&= \int \|[R^{\transp}_g(\boldsymbol{X}) \cdot \gamma_a, R^{\transp}_g(\boldsymbol{X}) \cdot \gamma_j]\|^2 \ d\mu(g) \\
	&&= \int \|[\boldsymbol{X} \cdot R_{gh}(\gamma_a), \boldsymbol{X} \cdot R_{gh}(\gamma_j)]\|^2 \ d\mu(g) \\
	&&= \int \|[\boldsymbol{X} \cdot R_g(\gamma_s), \boldsymbol{X} \cdot R_g(\gamma_t)]\|^2 \ d\mu(g)
	= \left[ \|[X_s,X_t]\|^2 \right]_G.
\end{eqnarray*}
Therefore,
\begin{eqnarray*}
	\left[ V_D \right]_G = \sum_{\genfrac{}{}{0pt}{}{a=8,9}{j=1,\ldots,7}} \left[ \|[X_a,X_j]\|^2 \right]_G 
	= \frac{14}{36} \sum_{s<t} \left[ \|[X_s,X_t]\|^2 \right]_G 
	= \frac{7}{18} [V]_G = \frac{7}{18} V.
\end{eqnarray*}

\subsection{Fermionic part}

The fermionic part of $H_D$, given in terms of Fock space operators
$\lambda_{\alpha' A} := \frac{1}{2}(\bt_{\alpha' A} + i\bt_{8+\alpha' \thinspace A})$,
$\alpha' = 1,\ldots,8$, is \cite{Hoppe3}
$$
	W_D = 2x_{jC}f_{CAB}( \delta_{\alpha' 8}\delta_{\beta' j} - \delta_{\alpha' j}\delta_{\beta' 8} )\lambda_{\alpha' A}\lambda_{\beta' B}^{\dagger}.
$$
With our choice of representation of the $\gamma$ matrices 
(see Appendix A of \cite{octonionic}), we find
$$
	W_D 
	= i \sum_{\rho=8,16} x_{jC} f_{CAB} \gamma^j_{\rho\beta} \bt_{\rho A} \bt_{\beta B}
	= i x_{jC} f_{CAB} [P\gamma^j]_{\alpha\beta} \bt_{\alpha A} \bt_{\beta B},
$$
where $P$ is a projection matrix s.t. $P_{8,8} = P_{16,16} = 1$ and zero otherwise.
Furthermore, one can verify that $P$ can be written as a product of 
three commuting projectors of the form $\frac{1}{2}(1 \pm E_\mu)$, $E_\mu^2=1$,
in the Clifford algebra:
\begin{equation} \label{P}
	P = \frac{1}{8}(1 - \gamma_1\gamma_2\gamma_3 I_7)(1 - \gamma_2\gamma_5\gamma_7 I_7)(1 - \gamma_3\gamma_6\gamma_7 I_7)
	= \frac{1}{8}(1 - CI_7),
\end{equation}
where $I_7 := \gamma_1\gamma_2\gamma_3\gamma_4\gamma_5\gamma_6\gamma_7$, and
$$
	C := \gamma^{123} + \gamma^{165} + \gamma^{246} + \gamma^{435} + \gamma^{147} + \gamma^{367} + \gamma^{257}
$$
defines an octonionic structure.
By choosing different signs for $E_\mu$ in the three projectors 
one obtains all $8 = 2^3$ projection matrices of that form.
Also note that $\gamma_1$, $\gamma_5$, and $\gamma_6$ share a particular
property in the expression \eqref{P}. 

The action \eqref{operator_action} yields
\begin{eqnarray*}
	\lefteqn{ [W_D]_G 
		= \int_{G} i (R^{\transp}_g(\boldsymbol{X}) \cdot \gamma_j)_C f_{CAB} [gP\gamma^jg^{-1}]_{\alpha\beta} \bt_{\alpha A} \bt_{\beta B}\ d\mu(g)
		}\\
	&&= \frac{1}{8} \sum_{p=1}^8 \int_{G} i (\boldsymbol{X} \cdot R_{gh_p}(\gamma_j))_C f_{CAB} [gh_pPh_p^{-1}g^{-1}R_{gh_p}(\gamma^j)]_{\alpha\beta} \bt_{\alpha A} \bt_{\beta B}\ d\mu(g),
\end{eqnarray*}
where we insert 8 different $h_p \in \Spin(7)$ such that
$R_{h_p}(\gamma_j) = \sigma_{p,j}\gamma_j \ \forall j$,
and $\sigma_{p,j} \in \{+,-\}$ are signs chosen so that
$\sum_p h_pPh_p^{-1} = 1$, e.g. according to the following table:
$$
	\begin{array}{c|ccccccc}
			p & \sigma_{p,1} & \sigma_{p,2} & \sigma_{p,3} & \sigma_{p,4} & \sigma_{p,5} & \sigma_{p,6} & \sigma_{p,7} \\
		\hline
		1	& + & + & + & + & + & + & + \\[-5pt]
		2	& + & + & + & - & + & - & + \\[-5pt]
		3	& + & + & + & - & - & + & + \\[-5pt]
		4	& + & + & + & + & - & - & + \\[-5pt]
		5	& - & + & + & - & + & + & + \\[-5pt]
		6	& - & + & + & + & + & - & + \\[-5pt]
		7	& - & + & + & + & - & + & + \\[-5pt]
		8	& - & + & + & - & - & - & + 
	\end{array}
$$
This is possible with $h_p \in \Spin(7)$ (and not only Pin(7)) 
because $\gamma_4$ does not appear explicitly in \eqref{P} except in $I_7$,
which with the choice of signs above is invariant, 
i.e. $h_p I_7 h_p^{-1} = R_{h_p}(\gamma_1) R_{h_p}(\gamma_2) \ldots R_{h_p}(\gamma_7) = I_7$.
Hence,
\begin{eqnarray*}
	\lefteqn{ [W_D]_G = }\\
	&&= \frac{1}{8} \sum_{j=1}^7 \int_{G} i (\boldsymbol{X} \cdot R_g(\gamma_j))_C f_{CAB} \left[ g \left({\textstyle \sum_p \sigma_{p,j}^2 h_pPh_p^{-1}}\right) g^{-1}R_g(\gamma^j) \right]_{\alpha\beta}\! \bt_{\alpha A} \bt_{\beta B}\thinspace d\mu(g) \\
	&&= \frac{1}{8} \sum_{j=1}^7 \int_{G} i (\boldsymbol{X} \cdot R_{gh'_j}(\gamma_j))_C f_{CAB} [R_{gh'_j}(\gamma^j)]_{\alpha\beta} \bt_{\alpha A} \bt_{\beta B}\ d\mu(g) \\
	&&= \frac{1}{4} \frac{7}{9} [H_F]_G,
\end{eqnarray*}
again using some appropriately chosen $h_j' \in \Spin(9)$.

\section{Result}

In total, we have
$$
	[H_D]_G = [-\Delta_{89}]_G + [V_{D}]_G + [W_{D}]_G 
	= -\frac{2}{9} \Delta_{\mathbb{R}^{9(N^2-1)}} + \frac{7}{2 \cdot 9} V + \frac{7}{4 \cdot 9} H_F.
$$
The relative coefficients of the terms of the resulting operator do not match those of $H$.
In fact, $[H_D]_G$ has a discrete spectrum on $\mathcal{H}_\textrm{phys}$
(contrary to $H$ whose spectrum covers the whole positive axis \cite{dWLN}).
This can be seen by rescaling the coordinates by $(\sqrt{7}/2)^{1/3}$, 
obtaining up to a constant
$$
	[H_D]_G \sim -\Delta + V + \kappa H_F = (1-\kappa)(-\Delta + V) + \kappa H
	\ge (1-\kappa)(-\Delta + V),
$$
with $\kappa = \sqrt{7}/4 < 1$. The observation follows since $H$ 
is a positive operator (by supersymmetry) and $-\Delta + V$ 
has a purely discrete spectrum \cite{Simon-Luscher}.

On the other hand, we can of course define a rescaled operator
$$
	H_D' := -\frac{9}{2}\Delta_{89} + \frac{2 \cdot 9}{7}V_D + \frac{4 \cdot 9}{7}W_D
$$
for which the average then is $[H_D']_{\Spin(9)} = H$.
Unlike $H_D$ which is positive due to supersymmetry, 
$H_D'$ has energies tending to negative infinity 
in certain regions of the $x'$ parameter space
(note that its $\mathpzc{h}$-eigenstates are still given by 
\eqref{eigenstates}, but with a rescaled frequency $S$). 
However, considering the action on $\Spin(9) \times \SU(N)$ invariant states 
$\Psi = U(g)\Psi$, we have
\begin{eqnarray}
	\lefteqn{ \langle \Psi, H_D' \Psi \rangle = \int \langle U(g^{-1})\Psi, H_D' U(g^{-1})\Psi \rangle \thinspace d\mu(g) } \nonumber \\
	&&= \left\langle \Psi, \int U(g) H_D' U(g)^{-1} d\mu(g) \Psi \right\rangle = \langle \Psi, [H_D'] \Psi \rangle \nonumber \\
	&&= \langle \Psi, H \Psi \rangle = \|\mathcal{Q}_\alpha \Psi\|^2 \geq 0. \label{quad_form_rel}
\end{eqnarray}
Hence, we conclude that these quadratic forms coincide on the 
subspace $\mathcal{H}_\textrm{inv}$ of invariant states, 
so that $H_D'$ and $H$ are actually the same operator on that 
subspace\footnote{The reader who is worried about
the unboundedness of the operators $H$ and $H_D'$ may consider
the dense subspace $\mathcal{H}_\textrm{inv} \cap C^\infty_0$, where
\eqref{quad_form_rel} makes perfect sense, and then conclude that
the Friedrichs extensions of $H_D'$ and $H$ on $\mathcal{H}_\textrm{inv}$
are equal.}.
Furthermore, because a zero-energy state of $H$ must be 
$\Spin(9)$ invariant \cite{Hasler-Hoppe} it is therefore sufficient to check that it is
annihilated by $H_D'$, i.e. that
$$
	\left( -7\Delta_{89} + 4\bar{z} \cdot S(x') z + 16W(x')\lambda \lambda^\dagger \right)\Psi(x',z) = 0 
	\quad \forall x'.
$$
Also note that the same holds for any linear combination, $(\alpha H + \beta H_D')\Psi = 0$.

\newpage

\subsubsection*{Acknowledgements}

We would like to thank 
V. Bach, M. Bj\"orklund and J.-B. Zuber for discussions,
as well as the Swedish Research Council and the
Marie Curie Training Network ENIGMA (contract MRNT-CT-2004-5652)
for financial support.

\end{document}